\documentclass[aps,preprint,a4paper,showpacs,showkeys,superscriptaddress]{revtex4-1}

\usepackage{latexsym}
\usepackage{amsmath,amssymb}
\usepackage{graphicx}
\usepackage{subfigure}
\usepackage{xcolor}
\usepackage{cancel}

\usepackage{hyperref}     
\hypersetup{colorlinks,%
  citecolor=blue,%
  linkcolor=cyan,%
  pdftex}
  
\begin{document}


\title{ On the thermodynamic origin of the initial radiation energy density in warm inflation}

\author{Yongwan Gim}%
\email[]{yongwan89@sogang.ac.kr}%
\affiliation{Department of Physics, Sogang University, 35 Baekbeom-ro, Mapo-gu, Seoul 04107, Korea}%
\affiliation{Research Institute for Basic Science, Sogang University, 35 Baekbeom-ro, Mapo-gu, Seoul 04107, Korea} %

\author{Wontae Kim}%
\email[]{wtkim@sogang.ac.kr}%
\affiliation{Department of Physics, Sogang University, 35 Baekbeom-ro, Mapo-gu, Seoul 04107, Korea}%

\date{\today}

\begin{abstract}
 In warm inflation scenarios, radiation always exists, so that
the radiation energy density is also assumed to be
finite when inflation starts. To find out the origin of the non-vanishing initial radiation energy density,
we revisit thermodynamic analysis for a warm inflation model
and then derive an effective Stefan-Boltzmann law which is commensurate with the temperature-dependent effective potential
by taking into account the non-vanishing trace of the total energy-momentum tensors.
The effective Stefan-Boltzmann law shows that
the zero energy density for radiation at the Grand Unification epoch increases
until the inflation starts and it becomes eventually finite at the initial stage of warm inflation.
By using the above effective Stefan-Boltzmann law, we
also study the cosmological scalar perturbation,
and obtain the sufficient radiation energy density in order for GUT baryogenesis at the end of inflation.
\end{abstract}


\maketitle


\section{Introduction}
\label{sec:intro}
In the big bang cosmology,
inflation is an elegant solution to the intriguing problems such as the horizon and flatness problems \cite{Starobinsky:1980te, Sato:1980yn, Guth:1980zm}.
It also generates the perturbations 
which are the origin of the spectrum of primordial gravitational waves \cite{Starobinsky:1979ty}, the cosmic microwave background (CMB) radiation
and the large scale structure of our universe \cite{Mukhanov:1981xt, Guth:1982ec, Hawking:1982cz, Starobinsky:1982ee, Bardeen:1983qw}.
The standard inflation, in particular, a chaotic inflation \cite{Linde:1981mu, Linde:1983gd}
is driven by scalar fields of the so-called inflaton.
 The inflationary expansion lays the universe in a supercooled phase,
and thereafter the universe is heated by assuming the reheating process.
In order to attain the explicit reheating process responsible for the graceful exit problem,
a wide variety of mechanisms of interest
have been studied
\cite{Abbott:1982hn, Albrecht:1982mp, Kofman:1994rk, Kofman:1997yn, Greene:1997ge, Allahverdi:2010xz, Amin:2014eta}.
However, in contrast to the assumption of the supercooled universe after inflation,
there has been another elegant way to approach this issue, that is, a warm inflation scenario without
reheating process \cite{Berera:1995ie, Berera:1996nv}.
The interactions of the inflaton and radiation are inevitable during inflation
via a damping term describing the decay rate of the inflaton into other fields,
and thus no large scale reheating is necessary,
where the curvature perturbations are generated
by a larger thermal fluctuation rather than a quantum fluctuation \cite{Berera:1995wh, Taylor:2000ze, Hall:2003zp}.

One of the most important ingredients
in the thermal history of the universe is
to determine the temperature at the end of inflationary regime, $i.e.$, the reheating temperature, $T_{\rm r}$, in the standard inflation models.
Even though the exact value of the reheating temperature has not yet been known,
the upper bound of the temperature has been considered as the scale of the grand unified theory (GUT),
 $T_{\rm r} \leq 10^{16}~{\rm GeV}$ \cite{Kolb1990},
 and the lower bound of the temperature has been constrained by the big bang nucleosynthesis as
 $T_{\rm r} \geq 10~{\rm Mev}$ \cite{Kawasaki:2000en, Hannestad:2004px}.
 Afterwards, another lower bound of the reheating temperature has been derived from the CMB data
based on the seven year Wilkinson
microwave anisotropies probe (WMAP7) data as
$T_{\rm r} \geq 6~{\rm Tev}$ \cite{Martin:2010kz}.
On the other hand, in the warm inflation model, the order of the temperature $T_{\rm end}$ at the end of inflation was obtained
as $T_{\rm end} \sim 10^{13} ~{\rm GeV}$ \cite{Hall:2003zp}.
And, in the presence of the non-minimal kinetic coupling model,
the temperature at the end of inflation was calculated
up to the uncertainties of the cosmological observations
as $5.01\times 10^{7} {\rm GeV} \le T_{\rm end} \le 2.11\times 10^{13}{\rm GeV}$
by the use of the formalism introduced in Ref. \cite{Mielczarek:2010ag}
with the data of Planck 2013
\cite{Goodarzi:2014fna}.

In the warm inflation scenario,
the radiation which is closely in thermal equilibrium always exists,
and thus
the initial radiation density is naturally assumed to be nonzero, $\rho_{\rm r}(t_{\rm i}) \neq 0$ \cite{Berera:2008ar},
which is
compatible with the Stefan-Boltzmann law of $\rho_r =3 \gamma T^4$
in the hot thermal bath at the initial point of inflation, $t=t_{\rm i}$.
On the other hand,
it was claimed that
the initial radiation energy density in thermal equilibrium with the thermal bath
is unjustified, and so it is required that $\rho_{\rm r}(t_{\rm i}) = 0$ when inflation starts,
following the spirit of the chaotic inflation scenario that the universe
should be created form a quantum fluctuation of the vacuum \cite{Bellini:2001ka}.
In this new scenario, the initial zero temperature is also increasing during inflation
based on the framework of the warm inflation scenario. Now, one might wonder
how to get the non-vanishing radiation energy density in warm inflation scenario.

In this work, within the framework of the warm inflation scenario along with thermodynamics,
we would like to investigate a possible way to get the initial non-vanishing radiation energy density from
the thermodynamic point of view.
If $\rho_{\rm r}(t_{\rm i}) \neq 0$, then it implies that the radiation energy density at the initial stage of inflation
should be thermodynamically originated before inflation.
For our purpose, we will assume that the radiation and inflaton are in thermal equilibrium
in order to use thermodynamic relations consistently, and
more importantly treat the inflaton as an equal footing with the radiation thermodynamically.
Consequently, we shall find that the usual Stefan-Boltzmann law which is only valid in
cases of the traceless energy-momentum tensor should be modified effectively
because the temperature-dependent effective potential gives rise to the non-vanishing trace of the
energy-momentum tensor.
In the end, the effective Stefan-Boltzmann law tells us that
the radiation energy density starts from zero with the GUT temperature as an initial condition of our universe,
and, subsequently, it increases and becomes finite, which eventually gives the adequate initial radiation energy density
for warm inflation.
By making use of the effective Stefan-Boltzmann law in the warm inflation scenario,
we also find a sufficient radiation energy density for the GUT baryogenesis
at the end of inflation.

In Sec.~\ref{sec:warminf},
we shall derive the effective Stefan-Boltzmann law from
the thermodynamic analysis by assuming that the total system which consists of
the inflaton and the radiation is closely in thermal
equilibrium.
In Sec.~\ref{sec:SR},
we present the slow-roll approximations in accord with the effective Stefan-Boltzmann law
defined in the previous section.
In Sec.~\ref{sec:perturb}, we obtain the explicit form of spectral index for the scalar perturbation
and the temperature at the end of inflation,
which will be numerically calculated by employing the data of Planck 2015 \cite{Ade:2015lrj, Ade:2015xua}.
The conclusion and discussion will be given
in Sec.~\ref{sec:ConDis}.

\section{effective Stefan-Boltzmann law in warm inflation}
\label{sec:warminf}

One of the most important ingredients in warm inflation is that the decreasing radiation energy density during inflation
is replenished in such a way that the energy of the inflaton field is transferred to that of radiation in virtue of dissipation.
It is worth noticing that only the radiation energy density is related to the temperature via
the Stefan-Boltzmann law in the standard warm inflation models.
As compared to this,
if one were to treat the inflaton and radiation on an equal footing in equilibrium,
then one would encounter generically non-vanishing trace of the total energy-momentum tensor due to the inflaton part
while the radiation part is still traceless.
Now, it should be emphasized that the usual
Stefan-Boltzmann law commonly rests upon the traceless condition of the energy-momentum tensor,
and thus we have to modify the Stefan-Boltzmann law in order
to incorporate the non-vanishing trace of the total energy-momentum tensor.

Let us start with the Helmholtz free energy defined by
$F = E-T S$, where $E$, $T$, and $S$ are the energy, temperature, and entropy
of a thermal system, respectively.
From the differential form of the Helmholtz free energy as $dF=dE-TdS-SdT$,
one can obtain the relation between the entropy $S$ and the Helmholtz free energy as $S=-\partial F/ \partial T$.
Using the Euler's relation of $E=TS-p {\rm V}$, one can also rewrite the Helmholtz free energy as
$F=-p {\rm V}$ where $p$ is the pressure and ${\rm V} $ is the volume of the system.
Then, it yields a relation for the entropy density
of $s= S/{\rm V}$ as
\begin{equation}\label{eq:spT}
s=\frac{\partial p}{\partial T}.
\end{equation}
On the other hand, the relevant energy-momentum tensor is assumed to be perfect fluid written as
$ T_{\mu\nu}=(\rho+p)u_\mu u_\nu +  g_{\mu\nu} p $,
where $u^\mu$ is the four-velocity of radiation flow satisfying $u^\mu u_\mu = -1$.
Assuming that the trace of the energy-momentum tensor is non-vanishing generically,
the trace relation is obtained as
$-\rho + 3p= T^\mu_\mu$ with the Euler's relation rewritten as
$ \rho +p=Ts$ where $\rho = E/{\rm V}$.
By eliminating the pressure in Eq. \eqref{eq:spT}, the differential equation for the energy density is obtained as
\begin{align}
T\frac{\partial \rho}{\partial T}-4 \rho &=T^\mu_\mu-T \frac{\partial T^\mu_\mu}{\partial T},
\end{align}
so that the effective Stefan-Boltzmann law to incorporate the non-vanishing trace of the energy-momentum tensor
can be obtained as
\begin{align}
\rho(T)=3 C_0 T^4 - \frac{1}{4}T^\mu_\mu - \frac{3}{4}T^4 \int^T \frac{1}{T^4}\frac{\partial T^\mu_\mu }{\partial T}dT,  \label{eq:SBrho}
\end{align}
and the pressure of
\begin{align}
p(T)= C_0 T^4 + \frac{1}{4}T^\mu_\mu - \frac{1}{4}T^4 \int^{T} \frac{1}{T^4}\frac{\partial T^\mu_\mu }{\partial T}dT, \label{eq:SBp}
\end{align}
where the integration constant $C_0$ can be fixed from an initial condition.
The relations \eqref{eq:SBrho} and \eqref{eq:SBp} naturally reduce  to the usual Stefan-Boltzmann law for the traceless case,
so that $C_0 =\gamma$. However, $C_0$ will be fixed for the case of the non-vanishing trace for our purpose later
by imposing a different boundary condition.
In fact, such a modified Stefan-Boltzmann law induced by conformal anomalies
had been applied to SU(3) lattice gauge theory
in particle physics in the Minkowski spacetime \cite{Boyd:1996bx} and the recent black hole physics
in connection with the information loss problem \cite{Gim:2015era}.

Now, from the cosmological point of view, let us assume that the total system of the early universe
consists of inflaton and radiation in thermal equilibrium.
Then  the total energy density $\rho_{\rm tot}$ and pressure $p_{\rm tot}$ are written
as \cite{Kolb1990}
\begin{align}
\rho_{\rm tot} &=\rho_\phi+\rho_{\rm r}= \frac{1}{2}\dot{\phi}^2 + V_{\rm eff}(\phi,T) +\rho_{\rm r}, \label{eq:rhotot} \\
p_{\rm tot} &=p_\phi+p_{\rm r} = \frac{1}{2}\dot{\phi}^2 - V_{\rm eff}(\phi,T) + p_{\rm r}, \label{eq:ptot}
\end{align}
where $\rho_{\rm r},~p_{\rm r}$ and $\rho_{\phi},~p_\phi$ denote the energy density and pressure of radiation
and inflaton, respectively. Specifically, the temperature dependent
effective potential $V_{\rm eff}$  for the inflaton is expressed by
\cite{Dolan:1973qd, Weinberg:1974hy, Linde:1978px}
\begin{equation}\label{eq:potential}
V_{\rm eff}(\phi, T)= - \gamma T^4 + \frac{1}{2}(\delta m_T)^2 \phi^2+V_0(\phi),
\end{equation}
where $\gamma=\pi^2 g_{*} /90$ and $g_{*}$ is an effective particle number.
$V_0(\phi)$ is the zero-temperature potential for the scalar field $\phi$, and $\delta m_T(\phi,T)$ denotes a thermal correction which will be neglected for simplicity along the lines of Ref. \cite{Hall:2003zp}.

The traceless condition for the radiation leads to
the equation of state as $p_{\rm r}=(1/3)\rho_{\rm r}$; however,
the trace for the total energy-momentum tensor
appears non-vanishing due to the effective potential for the inflaton as
\begin{align}
T^\mu_\mu =-\rho_{\rm tot}+ 3p_{\rm tot}  =  - 4V_{\rm eff}(\phi,T), \label{eq:trace}
\end{align}
where the kinetic energy is assumed to be very small as compared to the potential energy from now on.
By plugging Eq. \eqref{eq:trace} into Eqs. \eqref{eq:SBrho} and \eqref{eq:SBp}, the explicit forms of the pressure and energy density are obtained as
\begin{align}
\rho_{\rm tot} &= 12\gamma T^4 \ln \left( \frac{T_0}{T} \right) - \gamma T^4 +V_0(\phi), \label{eq:rhotot2}  \\
p_{\rm tot} &= 4\gamma T^4 \ln \left( \frac{T_0}{T} \right)+\gamma T^4 -V_0(\phi) \label{eq:ptot2}
\end{align}
by using the initial condition of $C_0= 4\gamma \ln T_0$
from the assumption that there exists only the inflaton field at the initial temperature of our universe $T_0$,
$i.e.$, $\rho_{{\rm tot}}(T_0)=\rho_\phi$ and $p_{{\rm tot}}(T_0)=p_\phi$.
Now, we take $T_0$  to be the GUT temperature as the maximum temperature of our universe
$T_0=T_{\rm GUT} = 10^{16} {\rm GeV}$, since
all perturbative interactions can be frozen out
and ineffective in maintaining or establishing thermal equilibrium for $T > 10^{16} {\rm GeV}$, and thus
the known interactions are not capable of thermalizing the universe at temperature greater than the GUT scale \cite{Kolb1990}.
Thus the energy density \eqref{eq:rhotot2} and pressure \eqref{eq:ptot2} are written as
\begin{align}
\rho_{\rm tot} = V_{\rm eff}+3\gamma T^4 \ln \left( \frac{T_{\rm GUT}}{T} \right)^4,~~
p_{\rm tot} =  -V_{\rm eff}+\gamma T^4 \ln \left( \frac{T_{\rm GUT}}{T} \right)^4. \label{eq:ptot3}
\end{align}
Comparing Eq. \eqref{eq:ptot3} with Eqs. \eqref{eq:rhotot} and \eqref{eq:ptot}, we can immediately find
the effective Stefan-Boltzmann law for the radiation as
\begin{align}
\rho_{\rm r} &= 3 \gamma T^4 \ln \left( \frac{T_{\rm GUT}}{T} \right)^4,\qquad
p_{\rm r} = \gamma T^4 \ln \left( \frac{T_{\rm GUT}}{T} \right)^4.   \label{eq:radrhop}
\end{align}
The traceless condition for the radiation is still met as $\rho_{\rm r} = 3p_{\rm r}$
which has been used not only in the warm inflation scenario \cite{Berera:1995ie, Hall:2003zp} but also
in the variety of cases of interest, for example, in the warm inflation model with the non minimal kinetic coupling \cite{Goodarzi:2014fna} and tachyon warm inflationary model \cite{Herrera:2006ck}.

It is interesting to note that
the energy density and pressure \eqref{eq:radrhop} for the radiation could be formally expressed
as the usual Stefan-Boltzmann law of $\rho_{\rm r} = 3 \gamma_{\rm eff} T^4$ and $p_{\rm r} = \gamma_{\rm eff} T^4$
when defining the temperature-dependent Stefan-Boltzmann constant
as $\gamma_{\rm eff}(T)=\gamma \ln \left( T_{\rm GUT}/T \right)^4$. So, the non-vanishing trace of the energy-momentum
tensor for the inflaton is of relevance to the modification of the Stefan-Boltzmann constant.
The physical consequence of this modification is that the radiation energy density vanishes at $T_{\rm GUT}$
and then it increases when the temperature of the universe decreases.
Subsequently, when inflation starts, it becomes finite
and gives the adequate energy density for radiation, and finally it
reaches $\rho_{\rm end}$ at $T_{\rm end}$ at the end of inflation, which will be elaborated in section IV.
This fact provides the reason why the standard warm inflation could assume
the finite radiation energy density when inflation starts.
 In essence, the radiation energy density could be thermodynamically created before inflation starts.

One might wonder why the form of the present Stefan-Boltzmann law
\eqref{eq:radrhop} is different from the previous one in Ref. \cite{Hall:2003zp}.
Apart from the additional consideration of the non-vanishing total energy-momentum tensor \eqref{eq:trace}
in the present thermodynamic analysis,
the other reason would stem from the different treatment of the temperature dependent term in
the finite temperature effective potential for the inflaton of $V_{\rm eff}(\phi, T)$.
From the total energy density \eqref{eq:rhotot} and  pressure \eqref{eq:ptot} with the effective potential \eqref{eq:potential}, the energy densities  and the pressures were identified with $\tilde{\rho}_{\phi}=\dot{\phi}^2/2 + V_0,~\tilde{p}_{\phi}=\dot{\phi}^2/2 - V_0$ and $\tilde{\rho}_{\rm r}=-\gamma T^4+\rho_{\rm r},~\tilde{p}_{\rm r}=\gamma T^4+p_{\rm r}$ in Ref. \cite{Hall:2003zp}.
In this case, the temperature-dependent term of $-\gamma T^4$ in $V_{\rm eff}(\phi, T)$ was incorporated into the radiation energy density and pressure rather than the inflaton energy density and pressure, so that the usual forms were obtained as
$\tilde{\rho}_{\rm r}=3 \gamma T^4,~\tilde{p}_{\rm r}=\gamma T^4$
by using the solution of $p_{\rm r}=0,~\rho_{\rm r} =4\gamma T^4$ obtained
from $\tilde{\rho}_{\rm r}+\tilde{p}_{\rm r}=T \tilde{s}_{\rm r},~\tilde{\rho}_{\rm r}=3\tilde{p}_{\rm r}$,
where $\tilde{s}_{\rm r}=\partial \tilde{p}_{\rm r}/\partial T$.
However, there is another way to identify the pressure such as
 $p_\phi = -V_{\rm eff}(\phi,T)$ with the effective potential \eqref{eq:potential} \cite{Kolb1990}.
 In this case,
the temperature dependent term in $V_{\rm eff}(\phi, T)$ was not included in the radiation part but it was incorporated in
the inflation part,
since it was originated from loop-corrections of inflaton field in thermal bath.
For the latter choice, Eq. \eqref{eq:radrhop} could be obtained.



\section{slow-roll approximations}
\label{sec:SR}
In this section, we consider the inflaton interacting with the radiation,
and thus the equations describing the system show how the energy lost by the inflaton through the damping force
is transferred to the radiation.
In the warm inflation model \cite{Hall:2003zp}, the energy conservation law, $\dot{\rho}_{\rm tot}+3H(\rho_{\rm tot}+p_{\rm tot})=0$,
can be separated into the inflaton and radiation parts as
\begin{align}
\dot{\rho_\phi}+3H(\rho_\phi+p_\phi) &= -\Gamma \dot{\phi}(t)^2, \label{eq:EOM1} \\
\dot{\rho_{\rm r}}+3H(\rho_{\rm r}+p_{\rm r}) &= \Gamma \dot{\phi}(t)^2,  \label{eq:EOM2}
\end{align}
where $H=\dot{a}/a$ denotes the Hubble parameter, and
$\Gamma \dot{\phi}^2$ is the friction term adopted phenomenologically
to describe the decay of the inflaton field and its energy transfers into the radiation bath.
And, the Friedmann equation for the evolution of the universe is also given as
\begin{equation}\label{eq:EOM3}
H^2-\frac{1}{3m_{\rm p}^2}\rho_{\rm tot}=0.
\end{equation}


Now, we exhibit slow-roll approximations which neglect terms of the highest order in time derivatives
in Eqs. \eqref{eq:EOM1}, \eqref{eq:EOM2}, and \eqref{eq:EOM3}
with the assumption that the inflaton field is dominant over the radiation field during the slow-roll warm inflation \cite{Hall:2003zp}, and thus we obtain
\begin{equation}\label{eq:SRapprox1}
\dot{\phi}^2 \ll V_{\rm eff} , \qquad \ddot{\phi} \ll \Gamma \dot{\phi} , \qquad
\dot{\rho_{\rm r}} \ll 4H \rho_{\rm r}, \qquad \rho_{\rm r} \ll \rho_\phi.
\end{equation}
By using a set of slow-roll parameters,
\begin{align}
\epsilon = \frac{m_{\rm p}^2}{2}\left( \frac{\partial_\phi V_{\rm eff}}{V_{\rm eff}}\right), \qquad \eta = m_{\rm p}^2\left( \frac{\partial_\phi^2 V_{\rm eff}}{V_{\rm eff}}\right),
\qquad \beta = m_{\rm p}^2\left( \frac{\partial_\phi V_{\rm eff}}{V_{\rm eff}}\right)\left( \frac{\partial_\phi \Gamma}{\Gamma} \right), \label{eq:SRparameters}
\end{align}
 the slow-roll approximations \eqref{eq:SRapprox1} can be summarized
as $\epsilon \ll r,~~\eta \ll r,~~ \beta \ll r $,
where $r$ is the ratio of the production rate of radiation, $\Gamma$, to the expansion rate, $3H$, defined as $r \equiv \Gamma/(3H)$.
Note that the slow-roll conditions are applied to the finite temperature effective potential \eqref{eq:potential}
rather than $V_0(\phi)$ which is contrast to the slow-roll conditions employed
in the standard warm inflation \cite{Berera:1995ie, Hall:2003zp}.
Neglecting several terms in Eqs. \eqref{eq:EOM1}, \eqref{eq:EOM2}, and \eqref{eq:EOM3}
based on the slow-roll approximations \eqref{eq:SRapprox1}, one can get the following equations,
\begin{align}
 3H r \dot{\phi} + \partial_\phi V_{\rm eff} &= 0, \label{eq:SReq1} \\
 3H (\rho_{\rm r}+p_{\rm r}) -\Gamma(\phi) \dot{\phi}^2 &= 0, \label{eq:SReq2} \\
H^2-\frac{1}{3m_{\rm p}^2}V_{\rm eff} &= 0,  \label{eq:SReq3}
\end{align}
where $r \gg 1$ in the warm inflationary regime.
Combining Eqs. \eqref{eq:SReq1} and \eqref{eq:SReq3}, one can rewrite Eq. \eqref{eq:SReq2} as
$\rho_{\rm r}+p_{\rm r} = m_{\rm p}( \partial_\phi V_{\rm eff})^2 /(\sqrt{3V_{\rm eff} }\Gamma)$,
and then obtain
\begin{equation}\label{eq:DiffT}
 4\gamma T^4 \ln \left( \frac{T_{\rm GUT}}{T} \right)^4  = \frac{m_{\rm p}( \partial_\phi V_{\rm eff})^2}{\sqrt{3V_{\rm eff}} \Gamma}
\end{equation}
by using the expressions for the energy density and pressure \eqref{eq:radrhop}.
%

Next, the number of e-folds during warm inflation is given as
\begin{align}
N_{\rm inf}=\int^{t_{\rm end}}_{t_{\rm HC}} H(t) dt
=\int^{\phi_{\rm HC}}_{\phi_{\rm end}} \frac{\Gamma \sqrt{V_{\rm eff}}}{\sqrt{3}m_{\rm p} \partial_\phi V_{\rm eff}}d\phi ,\label{eq:efold}
\end{align}
where $\phi_{\rm HC}$ and $\phi_{\rm end}$ are the values of the inflaton field corresponding  to
the horizon-crossing time $t_{\rm HC}$ and the end time of warm inflation $t_{\rm end}$, respectively.
Hereafter, in order to perform the specific calculations, we adopt the power-law potential $V_0$
and damping term $\Gamma$  \cite{Hall:2003zp} as
\begin{equation}\label{eq:Ex}
V_0(\phi)=\lambda \phi^n,  \qquad
\Gamma(\phi)=\Gamma_0 \left( \frac{\phi}{\phi_0}\right)^m,
\end{equation}
where the coefficients $\Gamma_0,~\phi_0$ and $\lambda$ are constants,
 and the power $n$ and $m$ are fixed as $n=2,~m=2$ for simplicity.
In this specific model, the number of e-folds \eqref{eq:efold} during inflation era is
finally written as
\begin{align}
N_{\rm inf} = \frac{\Gamma_0 (\lambda \phi_{\rm HC}^2-\gamma_{\rm HC} T^4_{\rm HC})^{\frac{3}{2}}}{6\sqrt{3} m_{\rm p} \lambda^2 \phi_0^2}  \label{eq:NSR}
\end{align}
by assuming that $\phi_{\rm end} \ll \phi_{\rm HC}$.


%
%
%
%

\section{Cosmological perturbations and temperature}
\label{sec:perturb}
In this section, let us determine the temperature bounds at end of inflation via cosmological perturbation.
The thermal fluctuations produce the power spectrum $P_\zeta$ for the comoving curvature $\zeta$ \cite{Hall:2003zp},
\begin{equation}\label{eq:P1}
P_\zeta =  \frac{\pi^{\frac{1}{2}}H^{\frac{5}{2}}\Gamma^{\frac{1}{2}}T}{2\dot{\phi}^2},
\end{equation}
and the power spectral index $n_{\rm s}$ for the scalar perturbation
is defined as $n_{\rm s}-1 = d\ln |P_\zeta |/d\ln k$,
which is calculated as
\begin{align}
n_{\rm s}-1 =\frac{5}{2H}\frac{d\ln H}{dt}+\frac{1}{2H}\frac{d\ln \Gamma}{dt}-\frac{2}{H}\frac{d\ln \dot{\phi}}{dt}+\frac{1}{H}\frac{d\ln T}{dt} \label{eq:ns1}
\end{align}
at the horizon crossing defined as $k=aH$,
where the relation $d\ln k \approx d\ln a = Hdt$ was employed.
Using the slow-roll equations \eqref{eq:SReq1}, \eqref{eq:SReq2}, and \eqref{eq:SReq3} with the entropy density,
we obtain the relations
\begin{equation}\label{eq:SRparameter2}
\frac{1}{H}\frac{d\ln H}{dt} = -\frac{1}{r} \epsilon, \qquad
\frac{1}{H}\frac{d\ln \Gamma}{dt} = -\frac{1}{r}\beta, \qquad
\frac{1}{H}\frac{d\ln \dot{\phi}}{dt} = \frac{1}{r}(\beta -\eta),
\end{equation}
and
\begin{equation}\label{eq:SRparameter3}
\frac{1}{H}\frac{d\ln T}{dt} = \frac{1}{4r}\left(1+\frac{1}{\ln\left(\frac{T_{\rm GUT}}{T}\right)^4-1} \right)(\epsilon+\beta-2\eta).
\end{equation}
From Eqs. \eqref{eq:SRparameter2} and \eqref{eq:SRparameter3},
we can express the spectral index in terms of slow-roll parameters \eqref{eq:SRparameters} as
\begin{equation}\label{eq:ns2}
n_{\rm s}-1 = \frac{3\eta}{2r}-\frac{9}{4r}(\epsilon+\beta)+\frac{1}{4r}\left(1+\frac{1}{\ln\left(\frac{T_{\rm GUT}}{T_{\rm HC}}\right)^4-1} \right)(\epsilon+\beta-2\eta),
\end{equation}
where $T_{\rm HC}$ is the temperature at the horizon crossing where the perturbation effectively occurs.
The spectral index \eqref{eq:ns2} is identical with the spectral index in Ref. \cite{Hall:2003zp}
except the last term  originated from the effective Stefan-Boltzmann law \eqref{eq:radrhop}.

\begin{figure}[hpb]
  \begin{center}
  \includegraphics[width=0.5\linewidth]{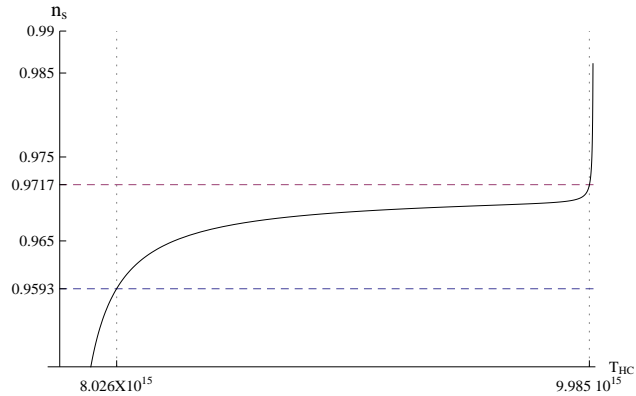}
  \end{center}
  \caption{The spectral index $n_{\rm s}$ vs the temperature at the horizon crossing $T_{\rm HC}$ is plotted
   such that the solid curve is the spectral index \eqref{eq:ns3}, where the number of e-folds and the GUT scale are fixed as $N_{\rm inf} = 60$ and $T_{\rm GUT} = 10^{16}{\rm GeV}$, and the two dashed lines show
    the range of the Planck 2015 data, $0.9593 \le n_{\rm s} \le 0.9717$.}
  \label{fig:nsTHC}
\end{figure}

With the help of numerical calculations,
the unknown parameters such as $\lambda,~\phi_{\rm HC},~\Gamma_0,~\phi_0$
are eliminated by combining Eqs. \eqref{eq:DiffT}, \eqref{eq:NSR} with Eq. \eqref{eq:ns2}.
After some
tedious calculations,
the spectral index is finally obtained as
\begin{equation}\label{eq:ns3}
n_{\rm s}-1=
\frac{1}{ N_{\rm inf}+N_{\rm inf}^2 \ln\left(\frac{T_{\rm GUT}}{T_{\rm HC}}\right)^4} - \frac{1}{12 N_{\rm inf}\left(1-\ln\left(\frac{T_{\rm GUT}}{T_{\rm HC}}\right)^4\right)} -\frac{7}{4 N_{\rm inf}},
\end{equation}
where
the number of e-folds is assumed to be $N_{\rm inf} = 60$ in order for solving the horizon problem.
As shown in Fig. \ref{fig:nsTHC}, the spectral index can respect the data of Planck 2015
when the temperature at the horizon crossing $T_{\rm HC}$ lies in the interval of
$8.026 \times 10^{15} {\rm GeV} \le T_{\rm HC} \le 9.985 \times 10^{15} {\rm GeV}$.

In order to evaluate the temperature
at the end of warm inflation,
we will use the procedure presented in Ref. \cite{Mielczarek:2010ag},
which was already applied to the non-minimal kinetic
coupling model \cite{Goodarzi:2014fna}.
By using Eq. \eqref{eq:SReq3}, the total number of e-folds $N_{\rm tot}$ from the scale at the horizon crossing $a_{\rm HC}$ to the scale at the present time $a_0$ is written as
\begin{align}\label{eq:N1}
N_{\rm tot} =\ln \left(\frac{a_0}{a_{\rm HC}} \right) =\ln \left(\frac{\sqrt{\lambda \phi_{\rm HC}^2-\gamma_{\rm HC}T_{\rm HC}^4 }}{\sqrt{3}k_0 m_{\rm p}}\right),
\end{align}
where the scale of the present time is fixed as
$a_0 = 1$ and the scale at the horizon crossing is given as $a_{\rm HC}=k_0/H(t_{\rm HC})$.
Next, the number of e-folds \eqref{eq:N1} can be divided into three parts composed of
inflationary regime $N_{\rm inf}= \ln (a_{\rm end}/a_{\rm HC})$,
 radiation-dominated era $N_{\rm rad}= \ln (a_{\rm rec}/a_{\rm end})$, and the time after recombination until now $N_{\rm 0}= \ln (a_{\rm 0}/a_{\rm rec})$ as
\begin{align}
N_{\rm tot}=N_{\rm 0}+N_{\rm rad}+N_{\rm inf}
=\ln \left(\frac{a_0}{a_{\rm rec}} \right)+\ln \left(\frac{a_{\rm rec}}{a_{\rm end}} \right)+\ln \left(\frac{a_{\rm end}}{a_{\rm HC}} \right), \label{eq:N2}
\end{align}
where $a_{\rm rec}$ and $a_{\rm end}$ are the scales at the recombination era and the end point of inflation, respectively.
The relation of
$T_{\rm rec}=(1+z_{\rm rec})T_{\rm CMB}$, where $z_{\rm rec}$ is the red-shift factor given as $1+z_{\rm rec} = a_0/a_{\rm rec}$,
indicates that the temperature diminishes
from the recombination era to present universe
due to the expansion of the universe. So,
the first term in Eq. \eqref{eq:N2} can be expressed as
\begin{equation}\label{eq:Nfirst}
N_0=\ln \left(\frac{a_0}{a_{\rm rec}}\right) = \ln \left(\frac{T_{\rm rec}}{T_{\rm CMB}}\right).
\end{equation}
For the radiation-dominated era in Eq. \eqref{eq:N2},
the adiabatic expansion of the universe is assumed as $dS=0$ \cite{Kolb1990},
so that $S= a_{\rm rec}^3 s_{\rm rec}=a_{\rm end}^3 s_{\rm end}$.
Then the number of e-folds can be rewritten in the radiation-dominated era $N_{\rm rad}$
as
\begin{equation}\label{eq:Nsecond}
N_{\rm rad} =\ln \left(\frac{a_{\rm rec}}{a_{\rm end}} \right)=\frac{1}{3}\ln \left(\frac{s_{\rm end}}{s_{\rm rec}}\right)
=\frac{1}{3} \ln \left(\frac{4\gamma_{\rm end} T_{\rm end}^3 \ln\left(\frac{T_{\rm GUT}}{T_{\rm end}}\right)^4}{4\gamma_{\rm rec}T_{\rm rec}^3}  \right),
\end{equation}
where the entropy density at the end of inflation is
$s_{\rm end}=4\gamma_{\rm end} T_{\rm end}^3 \ln\left(T_{\rm GUT} /T_{\rm end}\right)^4$
from Eqs. \eqref{eq:spT} and \eqref{eq:ptot3}.
By the way, $s_{\rm rec}=4\gamma_{\rm rec}T_{\rm rec}^3$ since the radiation only consists of photons without
the inflaton, so that the usual Stefan-Boltzmann law is used.
Plugging Eqs. \eqref{eq:N1}, \eqref{eq:Nfirst}, \eqref{eq:Nsecond} into  Eq. \eqref{eq:N2},
we get
\begin{equation}\label{eq:phiHE2}
\ln \left(\frac{\sqrt{\lambda \phi_{\rm HC}^2-\gamma_{\rm HC}T_{\rm HC}^4 }}{\sqrt{3}k_0 m_{\rm p}}\right) = N_{\rm inf}+
\frac{1}{3} \ln \left(\frac{\gamma_{\rm end} T_{\rm end}^3 \ln\left(\frac{T_{\rm GUT}}{T_{\rm end}}\right)^4}{\gamma_{\rm rec}T_{\rm CMB}^3}  \right).
\end{equation}
To determine $T_{\rm end}$,
we choose the effective particle number at the electroweak energy scale as $g_{\rm HC}=g_{\rm end}=106.75$
and at the recombination era as $g_{\rm  rec}=2$ \cite{Kolb1990}.
The temperature of CMB is known as $ T_{\rm CMB}=2.725 K$, and
the spectral index for $k_{0} = 0.05 {\rm Mpc}^{-1}$
is $n_{\rm s}=0.9655 \pm 0.0062$
from Planck 2015 \cite{Ade:2015lrj, Ade:2015xua}.
In the previous section, the temperature $T_{\rm HC}$ at the horizon crossing was already evaluated
as $8.026 \times 10^{15} {\rm GeV} \le T_{\rm HC} \le 9.985 \times 10^{15}$
by solving Eq. \eqref{eq:ns3}.
After all, from Eq. \eqref{eq:phiHE2},
the range of $T_{\rm end}$ is obtained as
\begin{equation}\label{eq:Tem1}
2.409\times 10^{13} ~{\rm GeV} \le T_{{\rm end}} \le 2.216 \times 10^{14} ~{\rm GeV},
\end{equation}
where this range lies below the well-known upper bound of the temperature of the universe to
avoid monopole proliferation
~\cite{Kolb1990}
and above the lower bounds in Refs.~\cite{Kawasaki:2000en, Hannestad:2004px, Martin:2010kz}.
In addition, the corresponding energy density for radiation is consequently
\begin{equation}
2.852 \times 10^{56} {\rm GeV}^4 \le \rho_{\rm end} \le 1.291 \times 10^{60} {\rm GeV}^4,
\end{equation}
which is a sufficient radiation energy density to accommodate
the GUT baryogenesis at the end of inflation \cite{Bellini:2001ka}.


\section{Conclusion and Discussion}
\label{sec:ConDis}
Motivated by the non-zero initial radiation energy density in warm inflation scenario,
we performed thermodynamic analysis for the warm inflation model
by using the definitions for the inflaton and radiation energy density presented in Ref. \cite{Kolb1990}.
And then we obtained the effective Stefan-Boltzmann law
to show that the zero radiation energy density \eqref{eq:radrhop} at
the Grand Unification epoch just prior to starting inflation became finite when inflation starts,
which gives the adequate radiation energy density for warm inflation.
By using the effective Stefan-Boltzmann law for the radiation energy density,
we studied the number of e-folds and the spectral index of the scalar perturbation under the slow-roll approximations
in the power-law potential and damping terms,
so that
the temperature \eqref{eq:Tem1} at the end of warm inflation was successfully calculated,
and it satisfies the upper bound lower than the GUT scale \cite{Kolb1990},
and lower bound of the big bang nucleosynthesis~\cite{Kawasaki:2000en, Hannestad:2004px}
by the CMB data \cite{Martin:2010kz}.
Additionally,  we confirmed that
a sufficient radiation energy density could be produced for GUT baryogenesis at the end of inflation \cite{Bellini:2001ka}.

As a matter of fact,
we have assumed the simplest setting described by a perfect fluid as a toy model;
however, the decay process causes
the deviation of equilibrium and perfectness of the radiation as well as
the inflaton field. So, there might be some deviations from this limit,
which leads to viscous dissipation and corresponding noise forces.
On general grounds, random sources and dissipative stresses are introduced via a shear stress tensor $\Pi_{\mu\nu}$
in the energy-momentum tensor, $ T_{\mu\nu}=(\rho+p)u_\mu u_\nu +  g_{\mu\nu} p + \Pi_{\mu\nu}$ \cite{Bastero-Gil:2014jsa}.
According to Landau's theory of random fluids \cite{Landau1975},
the dissipation is governed by constitutive relations for shear viscosity $\eta_{\rm s}$ and bulk viscosity $\eta_{\rm b}$
 while fluctuations are generated by Gaussian noise term $\Sigma_{\mu\nu}$.
In a comoving frame, the non-vanishing shear terms are written as
$\Pi_{\mu\nu}=-(\eta_{\rm s} \nabla_\mu u_\nu + \eta_{\rm s} \nabla_\nu u_\mu +(\eta_{\rm b}-2\eta_{\rm s}/3)\delta_{\mu\nu} \nabla_\kappa u^\kappa)-\Sigma_{\mu\nu}$.
In this case, the energy-momentum tensor
is obtained as $T^\mu _\mu =-\rho +3p + \Pi^\mu_\mu$.
Since the shear terms $\Pi_{\mu\nu}$ are the traceless part of the energy-momentum tensor,
the trace of the shear terms $\Pi^\mu_\mu$ automatically vanishes for the radiation and inflaton field, respectively.
So, the total
trace for the radiation and inflaton is simply coincident with our trace relation in this work,
and then the shear terms $\Pi_{\mu\nu}$ consequently
do not affect the form of the effective Stefan-Boltzmann laws \eqref{eq:SBrho} and \eqref{eq:SBp} thanks to the traceless property of shear terms.
On the other hand, the effects due to the shear terms $\Pi_{\mu\nu}$ play a role in the cosmological perturbation
 as seen from Ref. \cite{Bastero-Gil:2014jsa}.
To investigate specific changes due to the shear terms, we need to calculate the cosmological perturbation for the imperfect fluid with the effective Stefan-Boltzmann laws  \eqref{eq:SBrho} and \eqref{eq:SBp},
which appears to be a non-trivial task and becomes a harder problem.

Apart from the present phenomenological model of warm inflation,
one might be interested in other models derived
from the quantum field theory based on the first principles.
Let us discuss how our considerations would be applied to other warm inflation models.
For example, one can consider a model where the Higgs fields are coupled to left-handed fermions
$\psi_{\rm 1, L}$ and $\psi_{\rm 2, L}$ with U(1) charge $q$
as well as their right-handed counterparts,
$\psi_{\rm 1, R}$ and $\psi_{\rm 2, R}$, where one can take gauge singlets.
The finite temperature effective potential $V_{{\rm f}}$ from the fermion contribution \cite{Kapusta2006, Cline:1996mga}
takes the form of $V_{{\rm f}} \simeq \sum\limits_{i=1,2}[-(7\pi^2/180)T^4+ (m_i^2/12)T^2 + (m_i^4/(16\pi^2))(\ln(\mu^2/T^2)-c_f)]$
for $m_1,~m_2 \ll T$, where $\mu$ is the minimal subtraction renormalization scale and $c_f \simeq 2.635$.
Here, it is clear that the leading thermal inflaton mass corrections such as
$\delta m_{\rm T}(\phi,T)$ in the effective potential \eqref{eq:potential}
cancel out
by adding the contribution of both fermions \cite{Bastero-Gil:2016qru}.
Then by applying the effective Stefan-Boltzmann law \eqref{eq:SBrho} to this model
with the initial condition, $\rho_{{\rm tot}}=\rho_\phi$ at $T=T_{\rm GUT}$,
the energy density for radiation consisting of fermions is obtained as
$\rho_{\rm r}=(\pi^2/30)T^4 \ln(T_{\rm GUT}/T)^4 + (7\pi^2/60)T^4\ln(T_{\rm GUT}/T)^4
+(m_1^4+m_2^4)/(32 \pi^2)((3-2c_f)(1-(T/T_{\rm GUT})^4)+\ln(\mu/T)^4-(T/T_{\rm GUT})^4\ln(\mu/T_{\rm GUT})^4)$.
The behavior of this radiation density is very similar to that of the radiation density \eqref{eq:radrhop}
because of similar leading terms,
so that it starts from the zero radiation energy density and
subsequently generates a sufficient radiation energy density
to start the warm inflation.
Another example is the case where the potential is independent of the temperature
such as a renormalizable super potential as
$W=f(\Phi)+(g/2)\Phi X^2+(h/2)X Y^2$
with chiral superfields $\Phi,~X$ and $Y_i$, $i=1, ...,~N_{\rm Y}$ \cite{Berera:2002sp, BasteroGil:2010pb, BasteroGil:2012cm, Bastero-Gil:2014oga}.
In this model, the inflaton field corresponds to the scalar component of the chiral multiplet $\Phi$,
$\phi=\sqrt{2}\langle \Phi \rangle$,
and the associated scalar potential is $V(\phi)=|f'(\phi)^2|$,
which leads to supersymmetric breaking during inflation.
Using this model, one can show that the inflaton dissipates its energy into heavy fields and then it
decays into light degrees of freedom
in supersymmetric theories \cite{Bastero-Gil:2014oga}.
Since the trace $T^\mu_\mu$ for this model
is independent of the temperature, the third term of the effective Stefan-Boltzmann law \eqref{eq:SBrho} vanishes.
Hence, the energy density is obtained as $\rho(T)=3 C_0 T^4 - (1/4)T^\mu_\mu$.
In this case, the behavior of the radiation energy density is different from our
result \eqref{eq:radrhop} in the sense that the non-vanishing initial energy density should be assumed in the second model.
The effective Stefan-Boltzmann law can be applied to this model in principle, but it gives
the different behavior of the energy density as compared to that of the first model.
We hope that this issue will be elaborated in the near future since it deserves further attention.

%
%
%

\acknowledgments

We would like to Myungseok Eune and G. Tumurtushaa for exciting discussions.
W. Kim was supported by the National Research Foundation of Korea(NRF) grant funded by the Korea government(MSIP) (2014R1A2A1A11049571).

%



\bibliographystyle{JHEP}       

\bibliography{references}

\end{document}